\documentclass[12pt,a4paper]{article}
\usepackage{amsmath}
\usepackage{graphicx,color}
\usepackage{eps to pdf}
\color[rgb]{0,0,0}
\begin{document}
\sloppy

\begin{center}
EFFECT OF EXTERNAL MAGNETIC FIELD ON THE CO-EXISTENCE OF SUPERCONDUCTIVITY AND 
ANTIFERROMAGNETISM IN RARE EARTH NICKEL BOROCARBIDES $(RNi_{2}B_{2}C)$
\end{center}

\vspace{1.0cm}	

\centerline{$Salila$ $Das^1$,  $\forall Prakash$ $Chandra$ $Padhi^{2}$}
		\centerline{$^{1}$ Department of Physics, Berhampur University,
	Berhampur-760007, Odisha, India} 
    \centerline{$^{2}$ Bhabha Nagar (Lane-4),{Berhampur}-760004, Odisha, India}
\centerline{sd.phy@buodisha.edu.in}
 
\vspace{1.0cm}

\noindent \centerline{\Large Abstract}

In this paper we have studied the effect of external magnetic field in
the co-existing phase of superconducting and anti-ferromagnetism of rare earth nickel
borocarbides. The anti-ferromagnetism in these systems might have originated 
due to both localized 'f' electrons as well as itinerant electrons which 
are responsible for conduction. On the other hand, superconductivity is 
due to spin density wave, arising out of Fermi surface instability. The 
anti-ferromagnetism order is mostly influenced by
hybridization of the 'f' electron with the conduction electron. Here we have obtained 
the dependence of superconducting energy gap as well as staggered magnetic field on 
temperature T and energy $\epsilon_k$ in a framework based on mean 
field Hamiltonian using double time electron Green's function. We 
have shown in our calculation the effect of external magnetic 
field on superconducting and anti-ferromagnetic order parameter for 
$YNi_2B_2C$ in the presence of hybridization. The ratio of the calculated effective gap 
and $T_C$ is close to BCS value which agrees quite well with experimental results.
\newpage{}

\section{Introduction}

The rare earth nickel borocarbide $RNi_2B_2C$ (R = Y,Lu,Tm,Er,Ho,Dy) compounds are 
known for exhibiting  both superconductivity 
and long range anti-ferromagnetic order[1-2].
These compounds have two generally conflicting 
long range orders in an accessible temperature range and 
thus have attracted the attention of
many researchers in this field[1-4].  
They have layer structure consisting of $R-C$ sheets separated by 
$Ni_2B_2$ layers along the crystallographic 
c-axis[3] similar to that 
of high $T_c$ superconductors. 
In spite of the layered structure these compounds 
have isotropic normal state which are supported by band
structure calculation[5,6]. This result is supported experimentally 
by calculated resistivity  data in single crystal $YNi_2B_2C$ [7]. 
An explicit study on these compounds can disclose many effects on
the relationship between superconductivity and anti-ferromagnetism[8-9]. 
Hence, the large electron-electron coupling, high phonon frequencies and 
considerably large density of states at the Fermi surface, give evidence of the
phonon mediated BCS mechanism for superconductivity[6,9,10]. The experimental data shows an anisotropic[3]
nature of the superconducting state, which contradict with other observed data. In upper critical field measurement, 
$LuNi_2B_2C$ exhibits anisotropy when magnetic field
is applied parallel and perpendicular to c-axis which is not shown by Y- compounds[3,11].
Similarly, the experimental data on thermal conductivity[12] and 
photo-emission spectroscopic measurement [13] 
show large anisotropy[14] with the existence 
of node in the superconducting gap function. The specific heat, magnetization
and resistivity measurements point towards the s-wave symmetry
and the pairing is mediated by the electron-phonon interaction[15].

From the band structure calculation[5,6] of borocarbides, it is found that
one of the four bands crossing the Fermi level is a flat one indicates the
system being a correlated one. This flat band has the largest contribution 
to the density of states at Fermi surface suggesting it's dominance in the
formation of superconducting state [6,16]. On the other hand, the flat band hybridizes with the
conduction band formed out of other three bands (consisting of Y, B, C).  
There is a peak in the density of states at the Fermi 
level to which the d-band of nickel and bands of
Y(5d),B(2p)and C(2p) electrons [5,6,16] contribute.
Here we have extended the model as proposed by Fulde and others[17-20] to study the co-existence of 
superconductivity (SC) and anti-ferromagnetism (AFM) in presence of external magnetic field. 
In the present communication, we have included hybridization effect and the study of the effect of 
external magnetic field on the order parameters.The organization of the paper reflects: In section 2, the framework of
our model is briefly presented, where we have obtained the equation for 
superconducting and staggered magnetic field gap parameter by using double time Green
function of Zubarev type[21]. Section 3 
explains parameters for numerical calculation and discussion.
Finally section 4 provides the concluding remark.

\section{ Theoretical Framework }
\hspace{.3in} 

Superconductivity in the rare-earth nickel
borocarbides can be described by the usual BCS theory mediated by the phonon. The anti-ferromagnet 
might have originated from both the localized f-electrons and itinerant electrons responsible
for conduction. The itinerant electron which arises from spin density wave (SDW) due to  Fermi 
surface instability contributed to AFM, is also responsible for superconductivity.
Thus the localized magnetic moments due to staggered sub lattice magnetization co-exists
with superconductivity. Besides this, the electrons in a localized level hybridize 
with the conduction electrons near the Fermi level. Fulde et al [17] while describing the heavy fermion behaviour
for copper oxide superconductors, showed that the f-level of the rare earth atoms 
hybridizes with the conduction band. In  case of rare earth nickel borocarbides, 
the f-level of the rare earth atoms lies too far below the Fermi level to affect the electronic
properties of the system as seen from band structure calculations [5,6].
Due to hybridization, the sub-lattice AFM acquire the character of localized state of the flat 
d-band[18]. The order parameters corresponding to the AFM and the SC long range orders
have been calculated for rare earth nickel borocarbides[18,19].  
In our model Hamiltonian present below, we have included the effect of hybridization 
in presence of external magnetic field. The anti-ferromagnetic exchange leads to long range 
anti ferromagnetic order due to the spin alignment in Ni lattice site. This helps in 
dividing Ni lattice into two sub-lattices. If A and B denote two sub-lattice 
of Ni system, the Hamiltonian expressed in [17,18] can be extended to
\begin{eqnarray}
H=\sum_{\mathbf{k,}\sigma}\epsilon_{o}(\mathbf{k})(a^{\dag}_{\mathbf{k}\sigma}
b_{\mathbf{k}\sigma}+h.c.) + (\frac{h}{2})\sum_{\mathbf{k}\sigma}\sigma(a^{\dag}_{\mathbf{k}
\sigma}a_{\mathbf{k}\sigma}-b^{\dag}_{\mathbf{k}\sigma}b_{\mathbf{k}\sigma}) \nonumber \\
 +V\sum_{\mathbf{k}\sigma}(a^{\dag}_{\mathbf{k}\sigma}f_{1,\mathbf{k}\sigma}
+b^{\dag}_{\mathbf{k}\sigma}f_{2,\mathbf{k}\sigma}+h.c.)
-\Delta\sum_{\mathbf{k}}\left[(a^{\dag}_{\mathbf{k}\uparrow}
a^{\dag}_{-\mathbf{k}\downarrow}+h.c.)
 +(b^{\dag}_{\mathbf{k}\uparrow}b^{\dag}_{-\mathbf{k}\downarrow}+h.c.)\right]\nonumber\\ 
+\sum_{\mathbf{k}{\sigma}}(\epsilon_{f}+\sigma g\mu_{B}B)
\sum_{\mathbf{k}{\sigma,i=1,2}}f^{\dag}_{ik\sigma}f_{ik\sigma}+g\mu_{B}B\sum_{\mathbf{k}\sigma}\sigma(a^{\dag}_{\mathbf{k}
\sigma}a_{\mathbf{k}\sigma}+b^{\dag}_{\mathbf{k}\sigma}b_{\mathbf{k}\sigma})
\end{eqnarray}
where k, $\sigma$
$a^{\dag}_{\mathbf{k}\sigma}(a_{\mathbf{k}\sigma})$ and $b^{\dag}_{\mathbf{k}\sigma}
(b_{\mathbf{k}\sigma})$ are the momentum, spin, creation (annihilation) operators
of electrons belonging to the two sub lattices A and B respectively. The first term describes the hopping of 
conduction electrons between neighbouring sites of Ni and dispersion[17] is given by 

\begin{equation}
\epsilon_{o}(\mathbf{k})= -2t(\cos k_{x}+\cos k_{y})
\end{equation}
here t is the nearest neighbour hopping integral. The lattice constant has been set equal to unity.
The second term is due to staggered magnetic field h arising from Heisenberg exchange interaction
between the magnetic moments of neighborouring sites. This field acts on the Ni spins
and strongly reduces the charge fluctuation between different sites [18]. 'h' is the strength of the
sub lattice magnetisation (also anti-ferromagnetic order parameter) and is expressed [18] as
\begin{equation}
h=-\frac{1}{2}g\mu_{B}\sum_{\mathbf{k}\sigma}
	\left[<a_{\mathbf{k}\sigma}^{\dag}a_{\mathbf{k}\sigma}>
-<b_{\mathbf{k}\sigma}^{\dag}b_{\mathbf{k}\sigma}>\right]
\end{equation}
g and $\mu_{B}$ being the Lande g factor and Bohr magnetron, respectively.
The third term describes the effective hybridization.
$V$ is the hybridization interaction constant [17] and is taken to be independent of k.
For simplicity we have taken on-site 
hybridization (the localized electron 
belonging to sub-lattice 1 hybridize with the conduction electron of that sub-lattice 
alone and so on). The fourth term describes the attractive interaction of the 
charge carriers on the $Ni_2B_2$ planes leading to cooper pair formation. 
We have considered BCS type phonon mediated cooper pairing for superconductivity 
and intra sub-lattice cooper pairing is assumed to make the calculation simpler. The 
superconducting gap parameter $(\Delta)$ is defined as

\begin{equation}
\Delta=-\sum_{k}\tilde{V}_{k}\left(<a^{\dag}_{k\uparrow}a^{\dag}_{-k\downarrow}>
+<b^{\dag}_{k\uparrow}b^{\dag}_{-k\downarrow}>\right)
\end{equation}

$\tilde{V}_{k}$ is the strength of the attractive interaction between two electrons mediated by phonon. The intra f-electron Hamiltonian in presence of magnetic field is given by
 
\begin{equation}
\textit{H}_{f} =(\epsilon_{f}+\frac{1}{2}g\mu_{B}B)
\sum_{\mathbf{i,k_{i=1,2}}}f^{\dag}_{ik\uparrow}f_{ik\uparrow}+(\epsilon_{f}-
\frac{1}{2}g\mu_{B}B)
\sum_{\mathbf{i,k_{\mathbf{i=1,2}}}}f^{\dag}_{ik\downarrow}f_{ik\downarrow}
\end{equation}

$\epsilon_{f}$ is the dispersion less re normalized f-level
energy of the rare earth ions. $f^{\dag}_{ik\sigma}(f_{ik\sigma})$ is the creation
(annihilation) operator of the localized electron in the two
sub-latices (i=1,2). The external magnetic field acts on moments of rare earth and Ni element
which are parallel to each other. As a result, the external field splits the two field degenerate bands and 
new quasi particle energies are obtained. ( 
$\epsilon_F \rightarrow \epsilon_F \pm \frac{1}{2}g\mu_{B}B$ ). 
The Hamiltonian due to external magnetic field B on the Ni lattice is given by 
\begin{equation}
\textit{H}_{B} =\frac{1}{2}g\mu_{B}B[\sum_{\vec{k}}
(a^{\dag}_{\vec{k}\uparrow}a_{\vec{k}\uparrow}+
b^{\dag}_{\vec{k}\uparrow}b_{\vec{k}\uparrow})-
\sum_{\vec{k}}(a^{\dag}_{\vec{k}\downarrow}
a_{\vec{k}\downarrow}+b^{\dag}_{\vec{k}\downarrow}
b_{\vec{k}\downarrow})]
\end{equation}

\subsection{Superconducting gap} 
We study the Hamiltonian given in equation (1) with the help of Green's function technique using the equation of motion for 
the single particle green function of Zubarev type [21]. We have defined the
green functions $A_i(k,w)$, $B_i(k,w)$, $C_i(k,w)$, $D_i(k,w)$(i=1,6). We have
considered the following Green functions in our present calculation. 

\hspace{.3in}

\begin{eqnarray}
A_{1}(\mathbf{k},\omega) & = & \ll a_{\mathbf{k}\uparrow}
; a^{\dag}_{\mathbf{k}\uparrow}\gg_{\omega}\nonumber\\
A_{2}(\mathbf{k},\omega) & = & \ll a^{\dag}_{-\mathbf{k}\downarrow} ;
a^{\dag}_{\mathbf{k}\uparrow}\gg_{\omega}\nonumber\\
B_{1}(\mathbf{k},\omega) & = & \ll b_{\mathbf{k}\uparrow} ; b^{\dag}_{\mathbf{k}\uparrow}\gg_{\omega}\nonumber\\
B_{2}(\mathbf{k},\omega) & = & \ll b^{\dag}_{-\mathbf{k}\downarrow} ; b^{\dag}_{\mathbf{k}\uparrow}\gg_{\omega}\nonumber\\
C_{1}(\mathbf{k},\omega) & = & \ll a_{\mathbf{k}\downarrow} ; a^{\dag}_{\mathbf{k}\downarrow}\gg_{\omega}\nonumber\\
D_{1}(\mathbf{k},\omega) & = &\ll b_{\mathbf{k}\downarrow} ; b^{\dag}_{\mathbf{k}\downarrow}\gg_{\omega}\
\end{eqnarray}
For our convenience, we have dropped the k and w dependence of Green's function. 
The Fermi level is taken as zero $(\epsilon_{F}=0)$ and the
re-normalized localized f energy level is assumed to coincide with
the Fermi level for simplicity. The above Green's functions are evaluated by using equation 
of motion and commutation relation of the fermion operators
$a_{k\sigma}$, $b_{k\sigma}$, $f_{1k\sigma}$ .The coupled equations in $A_i(k,\omega)$,$B_i(k,\omega)$
$C_i(k,\omega)$, $D_i(k,\omega)$ for $i=1,2$ are solved and those equations can
be expressed in form of 

\begin{eqnarray}
A_{1,2}=\frac{\omega'}{4\pi}\left[\frac{\omega'(\omega'-
(\Delta-\frac{h}{2}))-V^{2}}
{\omega'^{4}-\omega'^{2}E_{1k}^2+V^{4}}\pm
\frac{\omega'(\omega'+(\Delta+\frac{h}{2}))-V^{2}}
{\omega'^{4}-\omega'^{2}E_{2k}^2+V^{4}}\right]
\end{eqnarray}
\begin{eqnarray}
B_{1,2}=\frac{\omega'}{4\pi}\left[\frac{\omega'(\omega'-
(\Delta+\frac{h}{2}))-V^{2}}
{\omega'^{4}-\omega'^{2}E_{2k}^2+V^{4}}\pm
\frac{\omega'(\omega'+(\Delta-\frac{h}{2}))-V^{2}}
{\omega'^{4}-\omega'^{2}E_{1k}^2+V^{4}}\right]
\end{eqnarray}
\begin{eqnarray}
C_{1}=\frac{\omega''}{4\pi}\left[\frac{\omega''(\omega''+
(\Delta-\frac{h}{2}))-V^{2}}
{\omega''^{4}-\omega''^{2}E_{1k}^2+V^{4}}+
\frac{\omega''(\omega''-(\Delta+\frac{h}{2}))-V^{2}}
{\omega''^{4}-\omega''^{2}E_{2k}^2+V^{4}}\right]
\end{eqnarray}
\begin{eqnarray}
D_{1}=\frac{\omega''}{4\pi}\left[\frac{\omega''(\omega''
+(\Delta+\frac{h}{2}))-V^{2}}
{\omega''^{4}-\omega''^{2}E_{2k}^2+V^{4}}+
\frac{\omega''(\omega''-(\Delta-\frac{h}{2}))-V^{2}}
{\omega''^{4}-\omega''^{2}E_{1k}^2+V^{4}}\right]
\end{eqnarray}
where

\begin{eqnarray}
E^{2}_{1k}=\epsilon^{2}_{1k}+2V^{2}\nonumber\\
E^{2}_{2k}=\epsilon^{2}_{2k}+2V^{2}\nonumber\\
\epsilon^{2}_{1k}=\epsilon^{2}_{0}(k)+(\Delta-\frac{h}{2})^{2}\nonumber\\
\epsilon^{2}_{2k}=\epsilon^{2}_{0}(k)+(\Delta+\frac{h}{2})^{2}\nonumber\\
\end{eqnarray}
and
\begin{eqnarray}
\omega'&=&\omega-\frac{1}{2}g\mu_{B}B\nonumber\\
h'&=&h+g\mu_{B}B\nonumber\\
h''&=&h-g\mu_{B}B
\end{eqnarray}

In the limit, when $h \rightarrow 0$, equation (12) reduces to BCS expression. The two band
vanishes in the absence of hybridization. The existence of four distinct bands in presence of 
hybridization indicates the co-existence of anti-ferromagnetism and superconductivity.
In the limiting condition when $\frac{h}{2} \rightarrow \Delta$ ,
$\epsilon_{1k} \rightarrow \pm \epsilon_0(k)$, superconductivity is suppressed by the anti-ferromagnetism. Similarly
$\epsilon_{2k} \rightarrow \pm[\epsilon_0^2(k) +2\Delta^2]^\frac{1}{2}$
shows superconductivity is enhanced by the anti-ferromagnetic order 
in these two bands. The poles of the Green's functions $A_{1,2}(w),B_{1,2}(w),C_{1}(w),D_{1}(w)$ are
obtained from
\begin{eqnarray}
\omega'^{4}-\omega'^{2}E_{1k}^2+V^4 =0\nonumber\\
\omega'^{4}-\omega'^{2}E_{2k}^2+V^4 =0
\end{eqnarray}
The solutions of the above equations are eight poles of the Green's functions
$A_{1,2}(k,\omega)$,$B_{1,2}(k,\omega)$. Solving the equation (14), we get 

\begin{eqnarray}
\omega'_{1\pm}=\pm\sqrt{\frac{1}{2}(E^{2}_{1k}+\sqrt{E^{4}_{1k}-4V^{4}})}\nonumber\\
\omega'_{2\pm}=\pm\sqrt{\frac{1}{2}(E^{2}_{1k}-\sqrt{E^{4}_{1k}-4V^{4}})}
\end{eqnarray}
and
\begin{eqnarray}
\omega'_{3\pm}=\pm\sqrt{\frac{1}{2}(E^{2}_{2k}+\sqrt{E^{4}_{2k}-4V^{4}})}\nonumber\\
\omega'_{4\pm}=\pm\sqrt{\frac{1}{2}(E^{2}_{2k}-\sqrt{E^{4}_{2k}-4V^{4}})}
\end{eqnarray}
These eight poles give eight quasi particle energy bands $\omega'_{i\pm}(i=1$ to $4)$. The superconducting
gap parameter ($\Delta$ ) in equation (4) can be explicitly written in terms of temperature dependent parameter
and its integral form which can be expressed as

\begin{eqnarray}
\Delta(T)= -\Sigma_{k}\widetilde{V}_{k}\left[
<a^{\dag}_{\mathbf{k}\uparrow}a^{\dag}_{-\mathbf{k}\downarrow}> +
<b^{\dag}_{\mathbf{k}\uparrow}b^{\dag}_{-\mathbf{k}\downarrow}>\right]
\end{eqnarray}

and 

\begin{equation}
\Delta(T)=-\widetilde{V}N(0)\int_{-\omega_D}^{+\omega_D}d\epsilon_0(\mathbf{k})\left[
<a^{\dag}_{\mathbf{k}\uparrow}a^{\dag}_{-\mathbf{k}\downarrow}> +
<b^{\dag}_{\mathbf{k}\uparrow}b^{\dag}_{-\mathbf{k}\downarrow}>\right]
\end{equation}

N(0) is the density of
state of the conduction electrons at the Fermi level
$\epsilon_{F}$ and $\omega_{D}$ is the Debye frequency. The limitation on the
k sum is due to the restriction of attractive interaction, which is effective 
with energies $\left|\epsilon_{1}-\epsilon_{2}\right|<\omega_{D}$.
In the weak coupling limit, the interaction potential $\widetilde{V}_{k}$ is     
\begin{flushleft}
\begin{eqnarray*}
\hspace{1.2in}\widetilde{V}_{k}=-V_{0}\hspace{1in}if\hspace{.3in} \left| \epsilon_{1}-\epsilon_{2}\right| < \omega_{D}\\
\hspace{1.2in}\hspace{.2in}=0\hspace{1in}\hspace{.8in}otherwise
\end{eqnarray*}
\end{flushleft}

Approximating the gap parameter to be
independent of $\mathbf{k}$, We can derive the equations for single particle co-relation function as [17,18] 
 
\begin{eqnarray}
<a^{\dag}_{\textbf{k}\uparrow}a^{\dag}_{-\textbf{k}\downarrow}>
+<b^{\dag}_{\textbf{k}\uparrow}b^{\dag}_{-\textbf{k}\downarrow}> =
i\lim_{\epsilon\rightarrow
0}\int\frac{d\omega}{(e^{\beta\omega}+1)}
\left[A_{2}B_{2}(\omega+i\epsilon)-A_{2}B_{2}(\omega-i\epsilon)\right]
\end{eqnarray}
Using equations (8) and (9) in equation (19), we obtain as
\begin{eqnarray}
<a^{\dag}_{\overrightarrow{\textbf{k}}\uparrow}a^{\dag}_{\overrightarrow{\textbf{-k}}\downarrow}>
+<b^{\dag}_{\overrightarrow{\textbf{k}}\uparrow}b^{\dag}_{\overrightarrow{\textbf{-k}}
\downarrow}> = \left[\frac{\Delta-\frac{h}{2}}{2\sqrt{E_{1k}^{4}-4V^{4}}}\left\{\omega'_{1}
\left(\frac{1}{e^{\beta\frac{\alpha}{2}}e^{\beta\omega'_{1}}+1}-\frac{1}{e^{\beta
\frac{\alpha}{2}}e^{-\beta\omega'_{1}}+1}\right)\right.\right.\nonumber\\
\left.-\omega'_{2}\left(\frac{1}{e^{\beta\frac{\alpha}{2}}e^{\beta\omega'_{2}}+1}
-\frac{1}{e^{\beta\frac{\alpha}{2}}e^{-\beta\omega'_{2}}+1}\right)\right\}\nonumber\\
+\frac{\Delta+\frac{h}{2}}{2\sqrt{E_{2k}^{4}-4V^{4}}}\left\{\omega'_{3}
\left(\frac{1}{e^{\beta\frac{\alpha}{2}}e^{\beta\omega'_{3}}+1}-\frac{1}{e^{\beta
\frac{\alpha}{2}}e^{-\beta\omega'_{3}}+1}\right)\right.\nonumber\\
\left.\left.-\omega'_{4}\left(\frac{1}{e^{\beta\frac{\alpha}{2}}e^{\beta\omega'_{4}}+1}-\frac{1}{e^{\beta
\frac{\alpha}{2}}e^{-\beta\omega'_{4}}+1}\right)\right\}\right]
\end{eqnarray}
The expression for the superconducting order parameter can be restructured  as

\begin{eqnarray}
\Delta(T)=-\tilde{V}N(0)\int^{+w_{D}}_{-w_{D}}d\epsilon_{o}(k)
[F_{1} (\overrightarrow{k,}T )+F_{2} (\overrightarrow{k,}T )]
\end{eqnarray}

where
\begin{eqnarray}
F_{1}\left(\overrightarrow{k,}T\right)=
\frac{\Delta-\frac{h}{2}}{2\sqrt{E_{1k}^4-4V^4}}\{ \omega'_{1}
(\frac{1}{e^{\beta\frac{\alpha}{2}}e^{\beta\omega'_{1}}+1}-\frac{1}{e^{\beta
\frac{\alpha}{2}}e^{-\beta\omega'_{1}}+1 }) \nonumber\\
 - \omega'_{2} (\frac{1}{e^{\beta\frac{\alpha}{2}}e^{\beta\omega'_{2}}+1}-\frac{1}{e^{\beta
\frac{\alpha}{2}}e^{-\beta\omega'_{2}}+1}) \}\nonumber\\
\end{eqnarray}
and
\begin{eqnarray}
F_{2}\left(\overrightarrow{k,}T\right)=
\frac{\Delta+\frac{h}{2}}{2\sqrt{E_{2k}^4-4V^4}}\{ \omega'_{3}
(\frac{1}{e^{\beta\frac{\alpha}{2}}e^{\beta\omega'_3}+1}
-\frac{1}{e^{\beta
\frac{\alpha}{2}}e^{-\beta\omega'_3}+1})\nonumber\\
-\omega'_4(\frac{1}{e^{\beta\frac{\alpha}{2}}e^{\beta\omega'_4}+1}-\frac{1}{e^{\beta
\frac{\alpha}{2}}e^{-\beta\omega'_4}+1})\}
\end{eqnarray}

\subsection{Staggered Magnetic Field h}
\hspace{.3in} The staggered magnetic field h as given in equation (3) is responsible for anti-ferromagnetism and
is assumed to be constant.The expression for staggered magnetic field strength in presence of external magnetic field 
is given as

\begin{eqnarray}
h=-\frac{1}{2}g\mu_{B}\sum_{\mathbf{k}}
[\{<a_{\mathbf{k}{\uparrow}}^{\dag}a_{\mathbf{k}{\uparrow}}>-<b_{\mathbf{k}{\uparrow}}^{\dag}b_{\mathbf{k}{\uparrow}}>\}
-\{<a_{\mathbf{k}{\downarrow}}^{\dag}a_{\mathbf{k}{\downarrow}}>-<b_{\mathbf{k}{\downarrow}}^{\dag}b_{\mathbf{k}{\downarrow}}>\}]
\end{eqnarray}
and
 
\begin{eqnarray}
h=-\frac{N(0)}{2}g\mu_{B}\int^{-\frac{W}{2}}_{+\frac{W}{2}}d\epsilon_{0}(k)
\left[\left\{<a_{\mathbf{k}{\uparrow}}^{\dag}a_{\mathbf{k}{\uparrow}}>-<b_{\mathbf{k}
{\uparrow}}^{\dag}b_{\mathbf{k}{\uparrow}}>\right\}-\left\{<a_{\mathbf{k}
{\downarrow}}^{\dag}a_{\mathbf{k}{\downarrow}}>-<b_{\mathbf{k}{\downarrow}}
^{\dag}b_{\mathbf{k}{\downarrow}}>\right\}\right]
\end{eqnarray}

Calculating the co-relation function
$<a_{\mathbf{k}{\uparrow}}^{\dag}a_{\mathbf{k}{\uparrow}}>$, 
$<a_{\mathbf{k}{\downarrow}}^{\dag}a_{\mathbf{k}{\downarrow}}>$,
$<b_{\mathbf{k}{\uparrow}}^{\dag}b_{\mathbf{k}{\uparrow}}>$
and $<b_{\mathbf{k} {\downarrow}}^{\dag}b_{\mathbf{k}{\downarrow}}>$ 
from the Green's functions$A_{1}(\mathbf{k},\omega)$, $B_{1}(\mathbf{k},\omega)$,
$C_{1}(\mathbf{k},\omega)$, and
$D_{1}(\mathbf{k},\omega)$, respectively, we can express the staggered magnetic field 'h' as

\begin{eqnarray}
h=-\frac{1}{2}g\mu_{B}\sum_{\vec{k}}\left[\left\{i\lim_{\epsilon\rightarrow0}\int\frac{d\omega}
{e^{\beta\omega}+1}\left(A_{1}B_{1}(\omega+i\epsilon)-A_{1}B_{1}(\omega-i\epsilon)
\right)\right\}\right.\nonumber\\
\left.-\left\{i\lim_{\epsilon\rightarrow0}\int\frac{d\omega}
{e^{\beta\omega}+1}\left(C_{1}D_{1}(\omega+i\epsilon)-C_{1}D_{1}(\omega-i\epsilon)
\right)\right\}\right]
\end{eqnarray}

\begin{eqnarray}
A_{1}B_{1}(\omega)=\ll
a^{\dag}_{\vec{k}\uparrow}a_{\vec{k}\uparrow}\gg-\ll
b^{\dag}_{\vec{k}\uparrow}b_{\vec{k}\uparrow}\gg=A_{1}(\omega)-B_{1}(\omega)
\end{eqnarray}

and
\begin{eqnarray}
C_{1}D_{1}(\omega)=\ll
a^{\dag}_{\vec{k}\downarrow}a_{\vec{k}\downarrow}\gg-\ll
b^{\dag}_{\vec{k}\downarrow}b_{\vec{k}\downarrow}\gg=C_{1}(\omega)-D_{1}(\omega)
\end{eqnarray}

Solving the above equations we can represent as

\begin{eqnarray}
A_1 (\omega)-B_1 (\omega) = -\left[\frac{\Delta-\frac{h}{2}}{2\sqrt{E_{1k}^{4}-4V^{4}}}\left\{\omega'_{1}
\left(\frac{1}{e^{\beta\frac{\alpha}{2}}e^{\beta\omega'_{1}}+1}-\frac{1}{e^{\beta
\frac{\alpha}{2}}e^{-\beta\omega'_{1}}+1}\right)\right.\right.\nonumber\\
\left.-\omega'_{2}\left(\frac{1}{e^{\beta\frac{\alpha}{2}}e^{\beta\omega'_{2}}+1}
-\frac{1}{e^{\beta\frac{\alpha}{2}}e^{-\beta\omega'_{2}}+1}\right)\right\}\nonumber\\
+\frac{\Delta+\frac{h}{2}}{2\sqrt{E_{2k}^{4}-4V^{4}}}\left\{\omega'_{3}
\left(\frac{1}{e^{\beta\frac{\alpha}{2}}e^{\beta\omega'_{3}}+1}-\frac{1}{e^{\beta
\frac{\alpha}{2}}e^{-\beta\omega'_{3}}+1}\right)\right.\nonumber\\
\left.\left.-\omega'_{4}\left(\frac{1}{e^{\beta\frac{\alpha}{2}}e^{\beta\omega'_{4}}+1}-\frac{1}{e^{\beta
\frac{\alpha}{2}}e^{-\beta\omega'_{4}}+1}\right)\right\}\right]
\end{eqnarray}

and

\begin{eqnarray}
C_1 (\omega)-D_1 (\omega) = [\frac{\Delta-\frac{h}{2}} {2\sqrt{E_{1k}^4-4V^4}}
\{ \omega''_1(\frac{1}{e^{-\beta\frac{\alpha}{2}} e^{\beta\omega''_1+1}} - \frac{1} {e^{-\beta\frac{\alpha}{2}}
{e^{-\beta\omega''_{1}}+1}})\nonumber\\
- \omega''_2 (\frac{1}{e^{-\beta\frac{\alpha}{2}}e^{\beta\omega''_{2}}+1}
- \frac{1}{e^{-\beta \frac{\alpha}{2}}e^{-\beta\omega''_{2}}+1}) \}\nonumber\\
-\frac{\Delta
+\frac{h}{2}}{2\sqrt{E_{1k}^4-4V^4}}
\{ \omega''_{3} (\frac{1}{e^{-\beta\frac{\alpha}{2}}e^{\beta\omega''_{3}}+1} 
-\frac{1}{e^{-\beta \frac{\alpha}{2}}e^{-\beta\omega''_{3}}+1})\nonumber\\
-\omega''_{4} (\frac{1}{e^{-\beta\frac{\alpha}{2}}e^{\beta\omega''_{4}}+1}-
\frac{1}{e^{-\beta \frac{\alpha}{2}}e^{-\beta\omega''_{4}}+1})\}]
\end{eqnarray}

where

\begin{eqnarray}
i\lim_{\epsilon\rightarrow0}\int\frac{d\omega}{e^{\beta\omega}+1}
\left[A_{1}B_{1}(\omega+i\epsilon)-A_{1}B_{1}(\omega-i\epsilon)\right]
=-\left[F_{1}\left(\overrightarrow{k,}T\right)-F_{2}\left(\overrightarrow{k,}
T\right)\right]
\end{eqnarray}

and

\begin{eqnarray}
i\lim_{\epsilon\rightarrow0}\int\frac{d\omega}{e^{\beta\omega}+1}
\left[C_{1}D_{1}(\omega+i\epsilon)-C_{1}D_{1}(\omega-i\epsilon)\right]
=\left[F_{3}\left(\overrightarrow{k,}T\right)-F_{4}\left(\overrightarrow{k,}
T\right)\right]
\end{eqnarray}
The extra term $F_{3}$ and $F_{4}$ are obtained due to application of magnetic field in
presence of hybridization. Where
\begin{eqnarray}
F_{3}\left(\overrightarrow{k,}T\right)
=\frac{\Delta-\frac{h}{2}}{2\sqrt{E_{1k}^4-4V^4}} \{ \omega''_{1}
(\frac{1}{e^{-\beta\frac{\alpha}{2}}e^{\beta\omega''_{1}}+1}-\frac{1}{e^{-\beta
\frac{\alpha}{2}}e^{-\beta\omega''_{1}}+1})\nonumber\\
-\omega''_{2}(\frac{1}{e^{-\beta\frac{\alpha}{2}}e^{\beta\omega''_{2}}+1}
-\frac{1}{e^{-\beta
\frac{\alpha}{2}}e^{-\beta\omega''_{2}}+1}) \}\nonumber\\
\end{eqnarray}
and
\begin{eqnarray}
F_{4}\left(\overrightarrow{k,}T\right)
=\frac{\Delta+\frac{h}{2}}{2\sqrt{E_{2k}^{4}-4V^{4}}}\{ \omega''_{3}
(\frac{1}{e^{-\beta\frac{\alpha}{2}}e^{\beta\omega''_{3}}+1}-\frac{1}{e^{-\beta
\frac{\alpha}{2}}e^{-\beta\omega''_{3}}+1}) \nonumber\\
-\omega''_{4}(\frac{1}{e^{-\beta\frac{\alpha}{2}}e^{\beta\omega''_{4}}+1}
-\frac{1}{e^{-\beta
\frac{\alpha}{2}}e^{-\beta\omega''_{4}}+1}) \}
\end{eqnarray}

Using equations (31) , (32) and (26) in equation (25), we can derive the $h(T)$ as
\begin{eqnarray*}
h(T)&=&-\frac{N(o)}{2}g\mu_{B}\int_{-\frac{W}{2}}^{+\frac{W}{2}}d\epsilon_{o}(k)\left\{-
\left[F_{1}(\overrightarrow{k},T)-F_{2}(\overrightarrow{k},T)\right]-
\left[F_{3}(\overrightarrow{k},T)-F_{4}(\overrightarrow{k},T)\right]\right\}\nonumber\\
\end{eqnarray*}
or
\begin{eqnarray}
h(T)&=&\frac{N(o)}{2}g\mu_{B}\int_{-\frac{W}{2}}^{+\frac{W}{2}}d\epsilon_{o}(k)\left\{
\left[F_{1}(\overrightarrow{k},T)-F_{2}(\overrightarrow{k},T)\right]
+\left[F_{3}(\overrightarrow{k},T)-F_{4}(\overrightarrow{k},T)\right]\right\}\hspace{0.5in}
\end{eqnarray}
Equations (21) and (35) are the final expressions for superconducting order parameter ($\Delta$)
and staggered anti-ferromagnetic order parameter(h). 
We have made all the parameters used in the above equation dimensionless by dividing them by 2t. 
The band width of the conduction band is taken as W=8t. 
Thus the dimensionless parameters are redefined as

\begin{eqnarray*}
\frac{\Delta(T)}{2t}=z &and& \frac{h(T)}{2t}=h
\end{eqnarray*}

\begin{eqnarray*}
\frac{\epsilon_{0}(k)}{2t}=x_{0},& \frac{\alpha}{2t}=\alpha,&
\frac{k_{B}T}{2t}=\theta
\end{eqnarray*}

Using the dimension-less quantity like phonon frequency
$\frac{\omega_{D}}{2t}=\tilde{\omega}_{D}$, the conduction
bandwidth $\frac{W}{2t}=\tilde{W}$, the strength of the
hybridization $\frac{V}{2(t)}=\tilde{V}$ and the 
coupling constants $N(0)V_{0}=\lambda_{1}$ (i.e,superconducting coupling constant),
$N(0)g\mu_{B}=\lambda_{2}$ (anti-ferromagnetic coupling constant),
the equations can be rewritten as 
\begin{eqnarray}
z=-\lambda_{1}\int^{+\tilde{\omega}_{D}}_{-\tilde{\omega}_{D}}dx_{0}(\mathbf{k})\left[F_{1}(x_{0},\theta)+F_{2}(x_{0},\theta)\right]
\end{eqnarray}
and
\begin{eqnarray}
h=\lambda_{2}\int^{\frac{\tilde{W}}{2}}_{-\frac{\tilde{W}}{2}}dx_{0}
\left[\left(F_{1}
(x_{0},\theta)-F_{2}(x_{0},\theta)\right)+\left(F_{3}(x_{0},\theta)
-F_{4}(x_{0},\theta)\right)\right]
\end{eqnarray} 
In the above equations $(\Delta)$ ,h  are self consistent equations. 
Thus in order to study these quantities with temperature, it is important to have the knowledge about the nature of the co-existing 
phase and these two equations have to be solved 
self-consistently. In our calculation the localised energy level is assumed to
coincide with the Fermi level i.e. $(\epsilon_{F}=0)$ as determined from band structure 
calculation [5].

\section{Results and Discussion}
In our calculation, we have obtained two parameters, 
superconducting gap function $(\Delta)$
and the staggered anti-ferromagnetic gap function (h) as given in equation (36) and (37) ,which are coupled to each
other. These equations are solved numerically and self-consistently 
in presence of magnetic field. We have used a standard set of parameters [5,6,19] ,like 
superconducting coupling $(\lambda_{1})= 0.111$, anti-ferromagnetic 
coupling $(\lambda_{2})=0.1598$, staggered magnetic field (h =0.001 to 0.005), 
conduction band width $(\tilde{\omega}_{D} = 1ev)$, 
temperature parameter ($\theta$ = 0.001 to 0.005) and hybridization strength 
(V = 0.002 to 0.0015). In an external magnetic field, the presence of hybridization in the Hamiltonian produces 
to extra terms ($F_3$ and $F_4$) in the equation for staggered magnetic field.
Fig-1. shows the temperature dependence of superconducting and anti-ferromagnetic order parameter for 
different external magnetic field. Superconducting gap increases with decreasing temperature and has an almost 
constant value around $\Theta_N$. In presence of external magnetic field the superconducting gap 
is suppressed in the pure phase due to intervene of magnetic moment of impurity preventing
the formation of spin up or spin down cooper pairs referring to as pair breaking . 
On the other hand superconducting gap 
is increased in the coexisting phase in an external magnetic field. 
When the external magnetic field is increased steadily from 0.0002(about 0.86T) to 0.0020(about 8.63T),
critical temperature parameter $\Theta_c$ decreases constantly from 0.0044 (equivalent to the temperature 12.69K ) 
to $\Theta_c$ = 0.0042 (about 12.12K). The calculated superconducting gap $\frac{2\Delta_0}{K_BT}$
increases from 1.46 to 1.62 agrees with the experimental results[9]. This is due to fact that 
when magnetic field is applied splitting of degenerate band takes place.Hence electron 
density of state increases which results
in increasing the probable availability of electron for cooper pairing. 
Thus superconducting order parameter increases towards 
lower temperature range below $T_N$ which agrees quite well with previous result [18].
From the graph 
it is observed that in presence of external 
magnetic field, Neel temperature 
get reduced from 5.74K to 4.14K whereas antiferromagnetic gap parameter $\frac{2\Delta_0}{K_BT_N}$ decreases
from 4.49 to 4.14.
Neel temperature around 5K, shows good agreement with our earlier 
calculation and experimental result [19,22].
                                                           
In figure-2. we have shown the effect of hybridization on superconducting 
and AFM order parameter. Both the order parameters get reduced with increase in hybridization. 
The hybridization plays an important role on z and h. In presence of hybridization 
of localized level with conduction band, the density of state at
Fermi level is reduced thereby reducing $T_c$ and $T_N$. The value of 
$\frac{2\Delta_0}{k_BT_c}$ increases from 1.46 to 1.62 and that of $\frac{2\Delta_0}{k_BT_N}$  increases 
from 3.49 to 4.04 with increasing V. Since the localized level 
lie at the Fermi level electrons from localized level can be transferred into Fermi 
level due to thermal fluctuation. This  contribute to cooper pairing in case of 
superconducting state and Neel ordering in case of anti ferromagnetic state. 
However, towards lower temperature range below Neel temperature, superconducting 
order parameter remains unaffected when hybridization is increased.
The critical temperature and Neel temperature decreases with increase of V. 
Here we have observed that the effect of hybridization is more drastic on the 
Neel ordering parameter as compared to superconducting order parameter 
in presence of external magnetic field. 
\newpage
\begin{figure}
\centering
\includegraphics{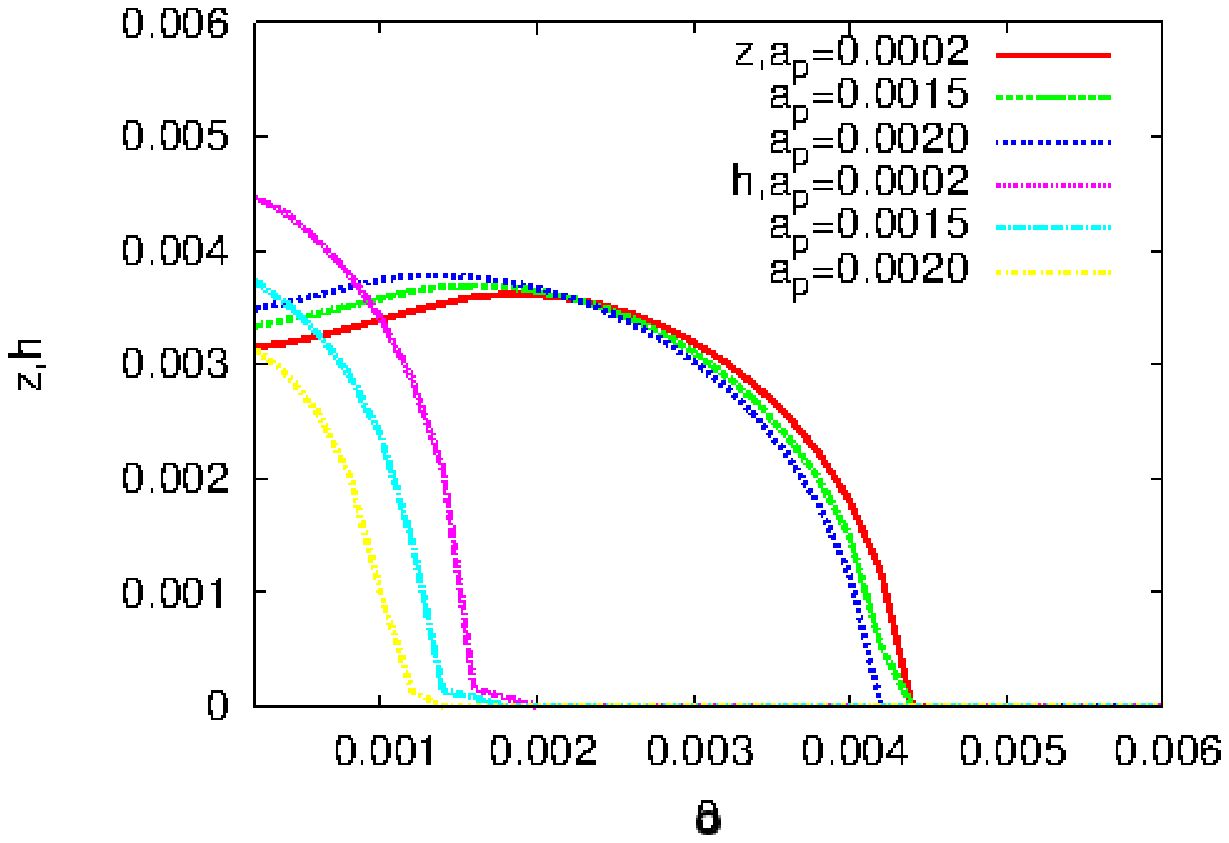}
\label{fig:fig1}
\end{figure}

\begin{figure}[htb]
\caption{ Sc gap(z) and AFM gap (h) vs. temperature 
for v=0.002 and various magnetic field. The superconducting 
coupling $\Lambda=0.111$, AFM coupling constant
$\Lambda_{1}=0.1598$, $w_{D} = 0.19$ }
\end{figure}

\begin{figure}
\centering
\includegraphics{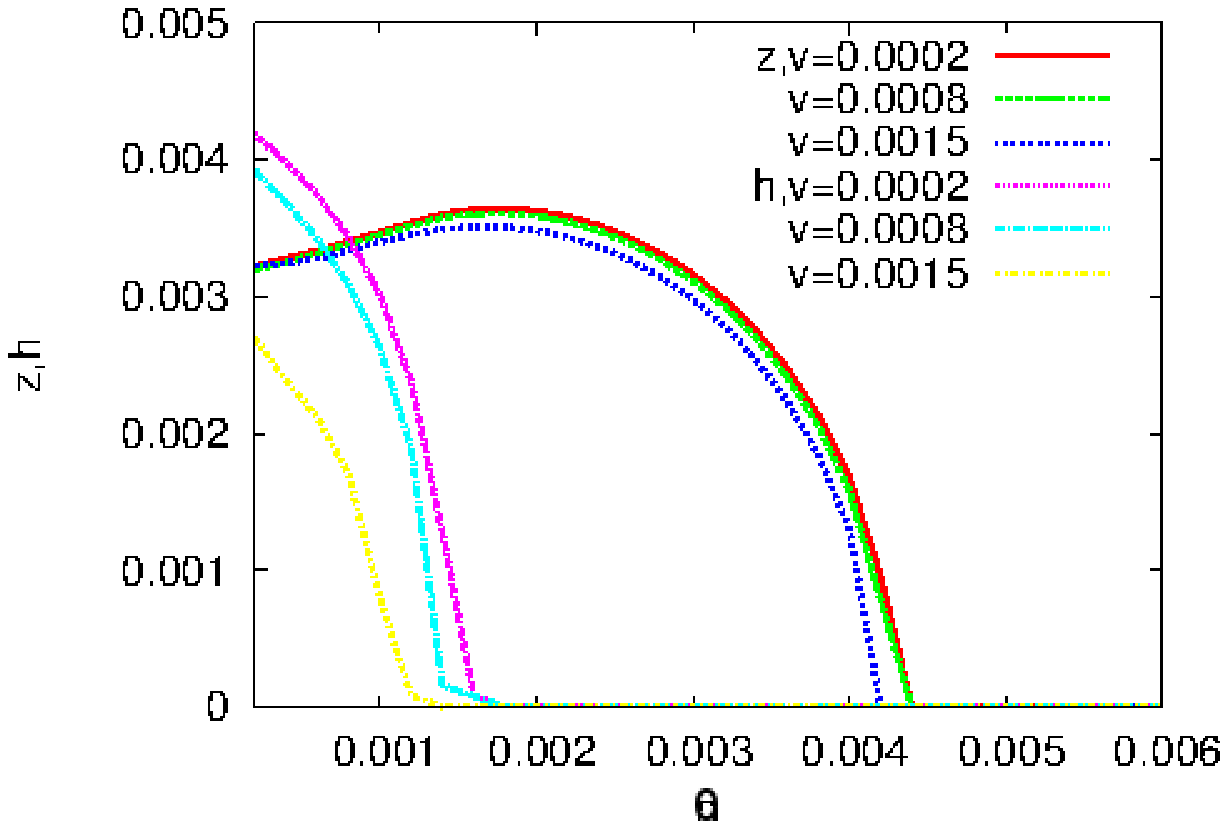}
\label{fig:fig3}
\end{figure}
\begin{figure}[htb]
\caption{Superconducting gap (z) and AFM (h) at 2 Tesla magnetic field for various hybridization parameter}
\end{figure}
\newpage
\section{Conclusion}
In this paper we have extended the model proposed by Panda et al [19] to take into account the effect of external field contribution 
in presence of hybridization. Along with other contributions, we have incorporated magnetic field contribution
in the model Hamiltonian [17,18,19] and obtained the expression for order parameters.
We have derived single particle 
Green's function using Zubarev's technique [21] and solved the superconducting and antiferromagnetic order parameters 
self-consistently with Fermi level at the middle 
of the conduction band. The variation of superconducting and anti-ferromagnetic
order parameter for different magnetic field is studied. The results are shown graphically which indicate 
similar trends as that of experimental 
result[22]. It is observed that hybridization 
reduces the long range magnetic order as well as Neel temperature but superconducting gap
parameter remain unaffected in the co-existing phase below $T_N$.
We have presented a simple model, which can account for the effect of external magnetic field on the 
co-existing states of superconducting and anti-ferromagnetism of Yttrium Nickel boro carbide in presence of hybridization. 
Similar studies for other compounds is being carried out.
\vspace{1.0cm}

\textbf{Acknowledgments}
 
 The authors gratefully acknowledge the useful discussions and technical inpute received from  Dr. Prafulla Kumar Panda, 
Reader in Physics, Utkal University, Odisha, Prof.R.N.Sahu, Prof.G.S.Tripathi, Berhampur University and Sri Santosh Kumar
Sahu, Brahma Nagar-2nd Lane, Berhampur. The authors also would like to acknowledge all those 
who have helped directly or indirectly for completion of this work. The co-author Mr.P.Ch.Padhi passed
away in the final stage of this work. My special acknowledgment for his dedication, endeavor and commitment
in initiating this piece of research and for his valuable suggestions to give it a final shape.

\newpage

\textbf{References} $\\$
\begin{enumerate}
\item P. C. Canfield, P. L. Gammel and D.J. Bishop, Phys.Today \textbf{51} (10),40 (1998).
\item R. Nagarajan, C. Mazumadar, Z. Hossian, S. K. Dhar et.al. Phys.Rev.Lett.\textbf{72},274 (1994).
\item Rare earth Transition Metal Borocarbides, superconductivity, Magnetic and Normal state prpoperties, 
edited by K.H.Muller and V.Naozhnyi (Kluwer Academic, Dordrecht, 2001) Rep.Prog. Phys.\textbf{64},943 (2001).
\item A. Amici, P. Talmeier and P. Fulde, Phys.Rev.Lett \textbf{64}, 943 (2001).
\item W. E. Picket and D. J. Singh Phys.Rev.Lett \textbf{72}, 3702 (1994)
\item L. F. Mattheiss Phys.RevB\textbf{49}, 13279 (1994).
\item K. D. D. Rathnayaka, A. K. Bhatnagar, A. Parasiris, D. G. Naugle, P. C. Canfield and B. K. Cho Phys.Rev.B\textbf{55},8506(1997).
\item K.Maki, P.Thalmeier and H.Won, Phys.Rev.B \textbf{65}, 140502 (2002);
\item I. K. Yanson, V. F. Fisun, A. G. M. Jansen, P. Wyder, P. C. Canfield, B. K. Cho, C. V. Tomy and D. M. Paul
Low.Temp.Phys \textbf{23},712(1997).
\item C. C. Thellwarth, P. Wavins and R. N. Shellon Phys.Rev.B \textbf{53},2579(1996).
\item V. Metlushko et.al. Phys.Rev.Lett. \textbf{79},1738 (1997).
\item K. Izawa et.al.Phys.Rev.Lett\textbf{89}, 137006 (2002).
\item K. Izawa et.al. Phys.Rev.Lett \textbf{86},1327 (2001).
\item S. A. Carter et.al. Phys.Rev.B\textbf{50}, 4216 (1994).
\item T. Yokoya et.al. Phys.Rev.Lett.\textbf{85}, 4952 (2000).
\item R. Coehoorn Physica C\textbf{228}, 331 (1994).
\item P. Fulde, V. Zenui, G. Zuricknagl Z.Phys B\textbf{92}, 133 (1993); P. Fulde, J Low Temp Phys \textbf{95}, 45 (1994).
\item S. N. Behera, B. N. Panda, G. C. Rout, P. Entel International Journal of Modern Physics B\textbf{15}, 2519 (2001); B. K. Sahu and 
B. N. Panda Modern Physics Letter B\textbf{24}, 2879 (2010).
\item B. K. Sahu and B. N. Panda Pramana \textbf{77}, 4, 715 (2011); B. K. Sahu and B. N. Panda Physica C 51335 (2015).
\item G. C. Rout, B. N. Panda and S. N. Behera Physics C\textbf{333}, 104 (2000).
\item D. N. Zubarev Sov Phys Usp \textbf{95}, 7 (1960).
\item D. L. Bashlakov et.al. Journal of Low temperature Physics\textbf{147} 314, 335 (2007).
\end{enumerate}
\end{document}